# Low temperature phase transformations in 4-cyano-4'-pentylbiphenyl (5CB) filled by multiwalled carbon nanotubes

N. Lebovka, [a,*] V. Melnyk,[b] Ye. Mamunya[c], G. Klishevich,[b]
A. Goncharuk,[a] N. Pivovarova[d]

[a]*Institute for Biocolloidal Chemistry, NAS of Ukraine, 42 Vernadsky Avenue, 03142 Kyiv, Ukraine, e-mail: lebovka@gmail.com (Lebovka N.)*

[b]*Institute of Physics, NAS of Ukraine, 46 Nauki Prosp, 03022 Kyiv, Ukraine, e-mail: melnyk@iop.kiev.ua (Melnyk V.)*

[c]*Institute of Macromolecular Chemistry, NAS of Ukraine, 48 Kharkivske chaussee, Kyiv 02160, Ukraine, e-mail: ymamunya@ukr.net (Mamunya Ye.)*

[d]*National University of Life and Environment Sciences of Ukraine, 15 Geroiv Oborony, Kyiv 03041, Ukraine, e-mail: natpiv@gmail.com (Pivovarova N.)*

*Corresponding author. Tel.: 380 44 4240378; Fax: 380 44 4248078
*E-mail address:* lebovka@gmail.com (N. Lebovka).

## Abstract

The effects of multiwalled carbon nanotubes (NTs) on low-temperature phase transformations in 5CB were studied by means of differential scanning calorimetry (DSC), low-temperature photoluminescence and measurements of electrical conductivity. The concentration of NTs was varied within 0-1% wt. The experimental data, obtained for pure 5CB by DSC and measurements of photoluminescence in the heating mode, evidenced the presence of two crystallization processes at $T$≈229 K and $T$≈262 K, which correspond to $C_1^a \rightarrow C_1^b$, and $C_1^b \rightarrow C_2$ phase transformations. Increase of temperature $T$ from 10 K до 229 K provoked the red shift of photoluminescence spectral band that was explained by flattening of 5CB molecule conformation. Moreover, the photoluminescence data allow to conclude that crystallisation at $T$≈229 K results in conformation transition to non-planar 5CB structure characteristic to ideal crystal. The non-planar conformations were dominating in nematic phase, i.e., at $T$>297 K. Electrical conductivity data for NTs-5CB composites revealed supplementary anomaly inside the stable crystalline phase $C_2$, identified earlier in the temperature range 229 K-296.8 K. It can reflect the influence of phase transformation of 5CB in interfacial layers on the transport of charge carriers between NTs.

## 1. Introduction

Nowadays, the studies of the properties of liquid crystalline composites, filled with nano-scale colloidal particles, present a great scientific and practical interest [1,2]. E.g., it was demonstrated that carbon nanotubes (NTs) with high aspect ratio (≥500-1000) can affect and improve the distinctive electro-physical, photonic and electro-optic characteristics of liquid crystals (LC) used in optical device and display applications [3]. Doping of LC by NTs allows reduction of the response time and driving voltage, as well as suppressing of the parasitic backflow and image sticking typical for LC cells [4–7]. Remarkable effects, such as electromechanical memory [8], super-elongation [9], anomalous electrokinetic dispersion [10] and ultra-low percolation thresholds [11–17] were also discovered.

The detailed study of optical and electrophysical properties of a nematic 5CB (4-cyano-4′-pentyl-biphenyl), filled with multilayer carbon NTs (0.0025-0.1 wt %), has shown the tendency to aggregation of NTs, formation of clusters with the fractal structure and percolation transition to a state with high electric conductivity [14–16]. Note that 5CB is a typical nematic LC, which exhibits crystalline, nematic and isotropic phases (Fig.1). The detailed studies have revealed also in 5CB the presence of rich solid crystalline polymorphism (SCP) dependent upon the thermal pre-history treatment [18]. Typically, SCP reflects the structure and short range order in the mesogenic phases; it does not totally exclude the conformational changes of the molecules [19]. The presence of SCP in 5CB was justified by Raman study [20,21] where the doublet in the C-N stretching region was observed and assigned to the presence of a polymorphic structure in 5CB [22]. The complicated photoluminesence spectral band was observed at low temperatures. It was explained by formation of different configurations of monomer and dimer structures of 5CB molecules [23–25]. The luminescence in nematic phase (near the room temperatures) was determined by the dimer structure, however, the mixture of monomer/dimer structures became dominating below the crystallization point and the fraction of dimer structure was decreasing with temperature.

Many properties of 5CB-NTs composites may be determined by the interfacial structures of 5CB near the surface of NTs. Note that analysis of the interfacial structures of 5CB on metal surfaces (Ag and Au) evidenced existence of the different binding schemes for 5 CB on metal surfaces [21]. Moreover, strong interaction between NTs and 5CB was justified by observation of micron-sized interfacial 5CB layers with irregular field of elastic stresses, and a complex structure of birefringence near NT clusters was observed [16].

The aim of the present work was the study of the effects of multiwalled carbon nanotubes (NTs) on low-temperatures phase transformations in 5CB. The experimental data were obtained using the methods of differential scanning calorimetry (DSC), low-temperature photoluminescence and measurements of electrical conductivity.

## 2. Materials and Methods

4-cyano-4′-pentyl-biphenyl (5CB) was purchased from Merck Ltd (Poole, GB). Its chemical composition is presented in Fig. 1. For pure 5CB, the crystalline–nematic transition temperature $T_{C \to N}$ is 296.8 K and nematic–isotropic transition temperature $T_{N \to I}$ is 308.4 K (the values are given by manufacture).

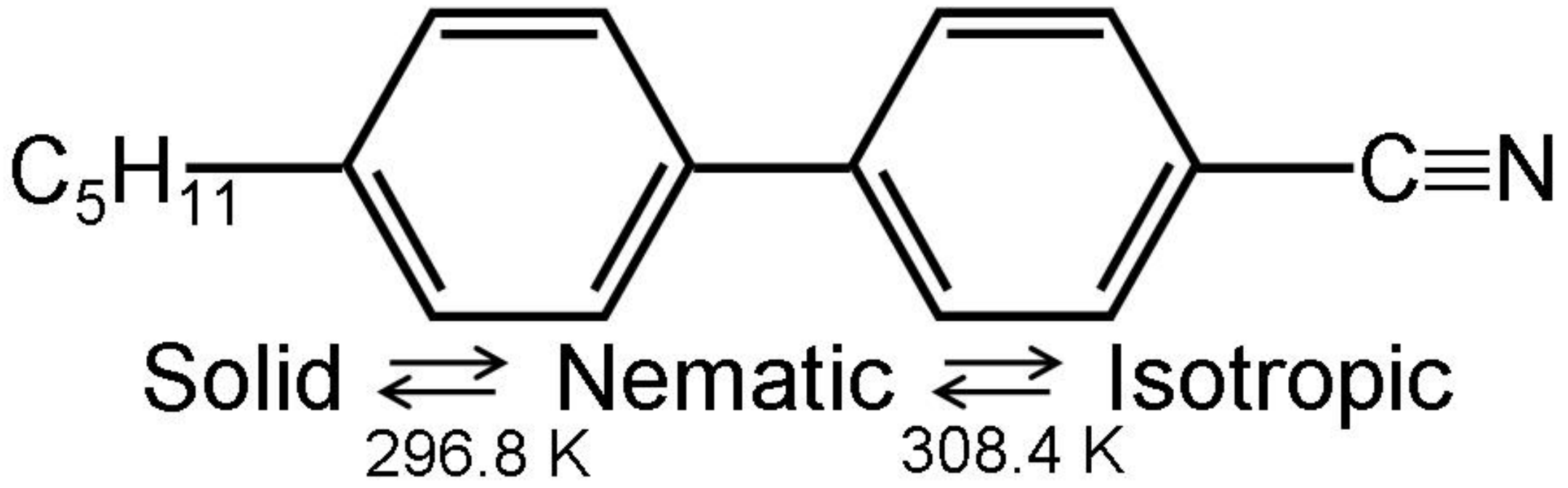

Fig. 1. Chemical structure of 5CB and phase transition temperatures.

The NTs were produced from ethylene by CVD method (TMSpetsmash Ltd., Kyiv, Ukraine) using FeAlMo as a catalyst [26]. The NTs were further treated by alkaline and acidic solutions and washed by distilled water until reaching of distilled water pH in the filtrate. The NTs had the outer diameter about 20-30 nm, while their length was approximately 5-10 microns.

The 5CB-NTs composites were prepared by addition of the relevant quantities of NTs to 5CB in the isotropic state ($T$=320 K). The weight concentration of NTs was varying within 0.1-1.0 %. The composites were sonicated for 5 min at the frequency of 22 kHz and the output power of 150 W using the ultrasonic disperser UZDN-2T (Ukrrospribor, Ukraine).

The calorimetric experiments were done using a differential-scanning calorimeter from TA Instruments, model DSC Q2000 equipped with a refrigerated cooling unit for controlled cooling and sub-ambient temperature operation. The measurements were carried out in the temperature range of 180–320 K in the cooling and heating modes, and the scanning rate was 5 K/min.

The electrical conductivity measurements were carried out in a conductometric cell, equipped with two horizontal platinum electrodes. The electrode diameter was 14 mm and the inter-electrode space was 0.5 mm. Before the measurements, the cell parts were washed in hexane and dried at 390 K. The electrical conductivity of samples was measured by the inductance, capacitance, and resistance (LCR) meter 819 (Instek, 12 Hz–100 kHz) under the fixed voltage $U$= 0.2 V and frequency of 0.5 kHz. High 0.5 kHz frequency was selected for avoidance of polarization effects on the electrodes and electrical field-induced asymmetric redistribution of NTs between the electrodes[27].

The conductivity measurements were carried out in the temperature range of 293–333 K in the heating and cooling modes with the scanning rate of 2 K/min. The temperature was stabilized using thermoelectric Peltier cooling modules (MT2-1.6-12 7S, 40x40 mm) in a home made thermostat and was recorded by a Teflon-coated K-type thermocouple (±0.1 K), connected to the data logger thermometer centre 309 (JDC Electronic SA, Switzerland).

The photoluminescence spectra were measured using spectrofluorimeter MPF-4 (Hitachi, Japan) and the exiting wavelength was 315 nm. The measurements were carried out in the temperature range of 10–300 K in the heating mode using a UTREX helium cryostat ("Sputnik”, Ukraine) equipped by the system of automatic temperature regulation, accurate to within 0.1 K. The sample of 5CB was placed in the quartz tubes with 5 mm inner diameter of and 25 mm height. Before the measurements it was rapidly cooled by direct immersion to the liquid helium.

## 3. Results and their Discussion

### *3.1. DSC data*

Figure 2 shows the DSC thermograms of 5CB-NTs composites measured during the cooling (a) and heating (b) modes. The samples were heated initially up to 310 K (to isotropic phase) and then were cooled at the rate of 5 K/min. The DSC peaks of isotropic→nematic transitions at $T_{I\to N}$≈308.5 K were narrow for pure 5CB and for low concentration of NTs ($C$=0.1%); however, they became broader at large concentration of NTs, $C$=1%. Strong supercooling of nematic phase and beginning of crystallization at $T_{N\to C}$≈260 K was observed for $C$=0-1%.. The increase of NT concentration resulted in increase of intensity and decrease of width of crystallization peaks. The low temperature crystallized phase was previously referred to as the metastable crystalline phase $C_1$[18].

Two crystallization (exothermic) peaks were observed for pure 5CB in the heating mode: the smaller was at $T$≈229 K and more intensive was at $T$≈262 K. The more intensive peak may be assigned to the crystallization transition observed earlier at $T$≈257 K [18]. It was attributed to transition of the metastable crystalline phase $C_1$ to solid state $C$, which melts at $T$≈296.8 K. The DCS data obtained in our work evidence that metastable phase $C_1$ may be subdivide into two crystalline phases $C_1^a$ and $C_1^b$. In presence of NTs, transition between the phases $C_1^a$ and $C_1^b$ became unobservable, and crystallization peaks, corresponding to transitions between the metastable $C_1$ and stable $C_2$ crystalline phases, became more distinct at high concentration of NTs, $C_{NT}$=1% (Fig. 2).

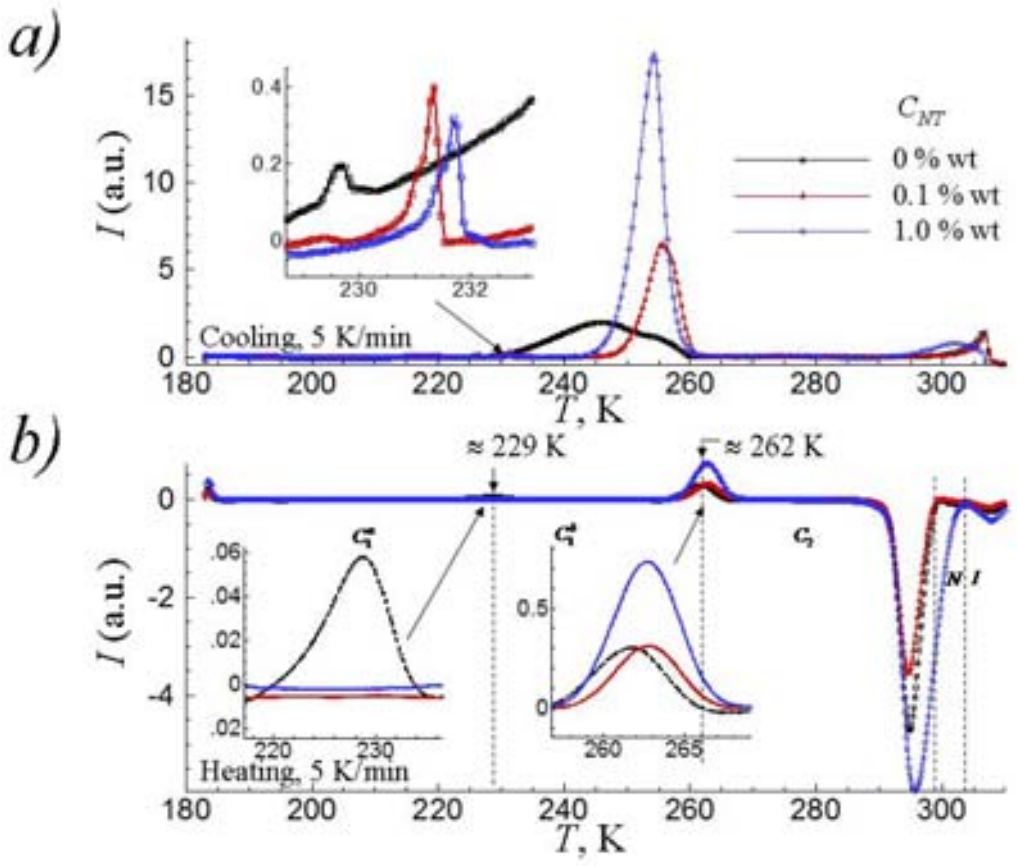

Fig. 2. DSC thermograms of 5CB-NTs composites at different concentrations of NTs, $C_{NT}$, measured during the cooling (a) and heating (b) modes. Inserts show enlarged portions of thermograms.

### 3.2. Photoluminescence spectra

Figure 3 present the photoluminescence spectra of pure 5CB and NTs-5CB composite ($C_{NT}$=1%) at two different temperatures, $T$=10 K and $T$=297 K.

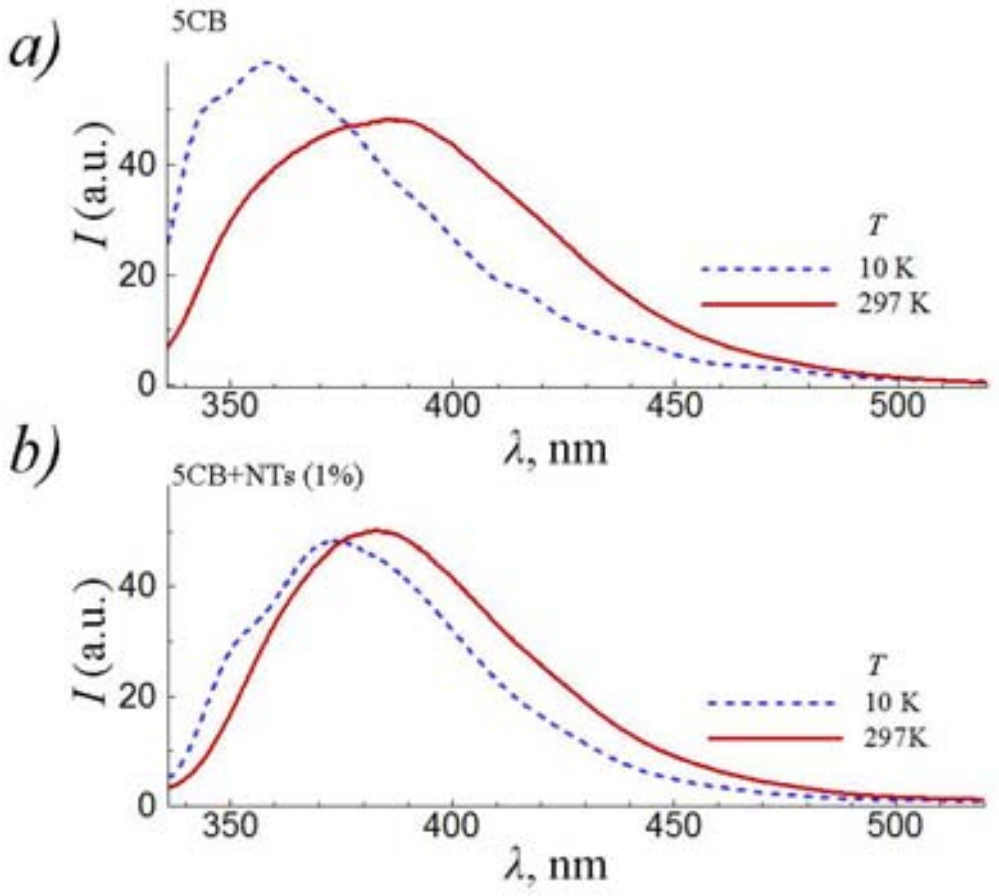

Fig. 3. Photoluminescence spectra of pure 5CB and NTs-5CB composite ($C_{NT}$=1%) at two different temperatures, $T$=10 K and $T$=297 K.

Increase of temperature $T$ from 10 K до 297 K resulted in the long-wave (red) shift of band maximum and induced changes in the shape of spectral band for both pure 5CB and NTs-5CB composite. These results can be explained by conformational changes in the structure of 5CB molecules with temperature increase[25]. The alkyl chains of 5CB in the low-temperature crystalline phase are in extended conformation. They are oriented practically perpendicular to the plane of the nearest benzene ring (the torsion angle is ≈90°) and dihedral angle is ≈32[28]. Owing to the high polarity of 5CB molecules and strong dipole-dipole interactions between them, they can form dimer structures with planar conformation of 5CB. These structures were shown to be dominating in nematic phase of 5CB [29]. The energy gap between ground state $S_0$ and first excited $S_1$ conformations decreases decreases with increase of the fraction of molecules with planar conformation [30], that's why the experimentally observed long-wave shift may reflect flattening of 5CB molecules with temperature increase. Note that previous investigation of photoluminescence of 5CB revealed the presence of several types of crystal modifications with different monomer and dimer conformers below 160K [25]. The value of red shift, corresponding to the temperature increase from 10 K до 297 K, was noticeably larger in pure 5CB (27.2 nm) than in 1% NTs-5CB composite (10.47 nm). This effect may reflect the presence of strong perturbation of the crystalline structure and conformational state of 5CB molecules in the presence of 1% of NTs.

The more detailed analysis revealed complex temperature dependencies of the band maximum $\lambda$ and its half-width $\Delta\lambda$ in photoluminescence spectra of pure 5CB and NTs-5CB composite (Fig. 4). The values of $\lambda$ and $\Delta\lambda$ were symbatically changing with temperature. For pure 5CB, the values of $\lambda$ and $\Delta\lambda$ initially increased within the temperature range of 10-230 K. It may be explained by flattening of 5CB molecules with temperature increase [30]. The evident crystal–crystal transformation at $T \approx$ 80K (it was denoted as transformation between the crystalline phases $C_o^a$ and $C_o^b$ in Fig. 4a) was accompanied by the total loss of fine structure in 5CB photoluminescence spectrum and disappearance of the spectral band at 343 nm [25]. From the other side, the saturation observed in $\lambda(T)$ and $\Delta\lambda(T)$ dependencies above $T \approx$ 140 K corresponded to appearance of a new spectral band at 424 nm [25]. The crystal–crystal transformation at $T \approx$ 140 K (it was denoted as transition between the crystalline phases $C_o^b$ and $C_1^a$ in Fig. 4a) was previously explained by formation of the overlapping dimers of 5CB.

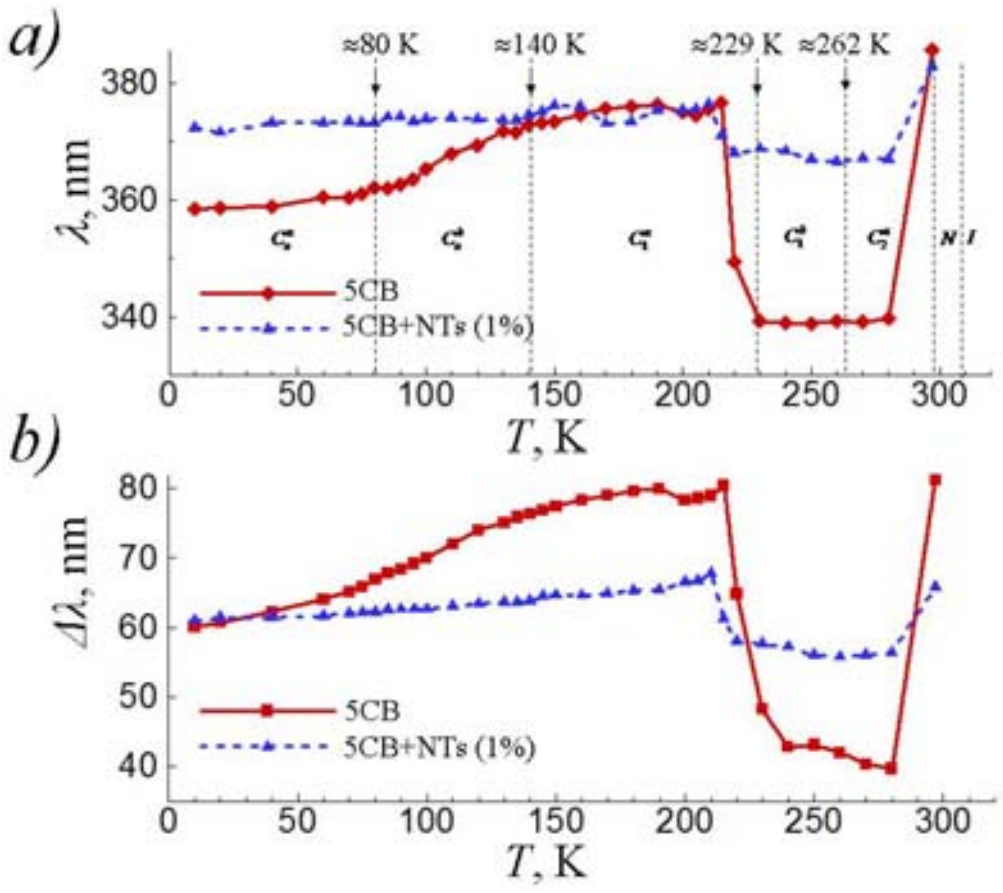

Fig. 4. Temperature dependencies of band maximum λ and its half-width Δλ in photoluminescence spectra of pure 5CB and NTs-5CB composite ($C_{NT}$=1%).

The noticeable dips in temperature dependences of $\lambda$ and $\Delta\lambda$ were observed between 230 K and ≈295 K. The sharp transition at $T \approx 230$ K correlated with transition between the metastable crystalline phases $C_1^a$ and $C_1^b$ observed from DSC data (Fig. 2b) at 229 K. It is reasonable to assume that observed $C_1^a \rightarrow C_1^b$ transition can reflect cooperative low-temperature transformations inside the metastable phase. The photoluminescence data did not reveal $C_1^b \rightarrow C_2$ transition at $T$=262 K (Fig. 4) that was revealed from DSC data. Possibly, it reflected different thermal prehistories of the samples in DSC and photoluminescence measurements. The $C_1^a \rightarrow C_1^b$ transformation was less observable in photoluminescence data (Fig. 4) and practically invisible in DSC data in the presence of NTs (Fig.2b). This fact can be explained by partial elimination of the metastable state of 5CB in the presence of NTs. Acceleration of crystallisation processes in 5CB in the presence of NTs supports this conclusion (Fig. 2a).

### *3.2. Electrical conductivity*

For better insight into the phase transformations in the temperature range between the crystalline phases $C_1$, $C_2$ and nematic phase *N*, the supplementary measurements of electrical conductivity $\sigma$ were done for the NTs-5CB composites with concentration of NTs above the percolation threshold ($C_{NT}$=0.1-1%). The previous experimental data for NTs-5CB composites evidenced an abrupt growth of $\sigma$ by several order of magnitude with increase of NT concentration and percolation transition from non-conducting to conducting state was observed at $C_{NT}$≥0.025-0.1% [15,16]. Moreover, the hysteretic temperature behaviour of $\sigma$ was observed. It was explained by spatial rearrangement of NTs inside LC. Electrical conductivity near the percolation threshold can be rather sensible to structural organisation of NT networks, the interfacial structures of 5CB near the surface of NTs, and phase state of 5CB in bulk. So, electrical conductivity may be a sensible tool for detection of the different kinds of phase transformations.

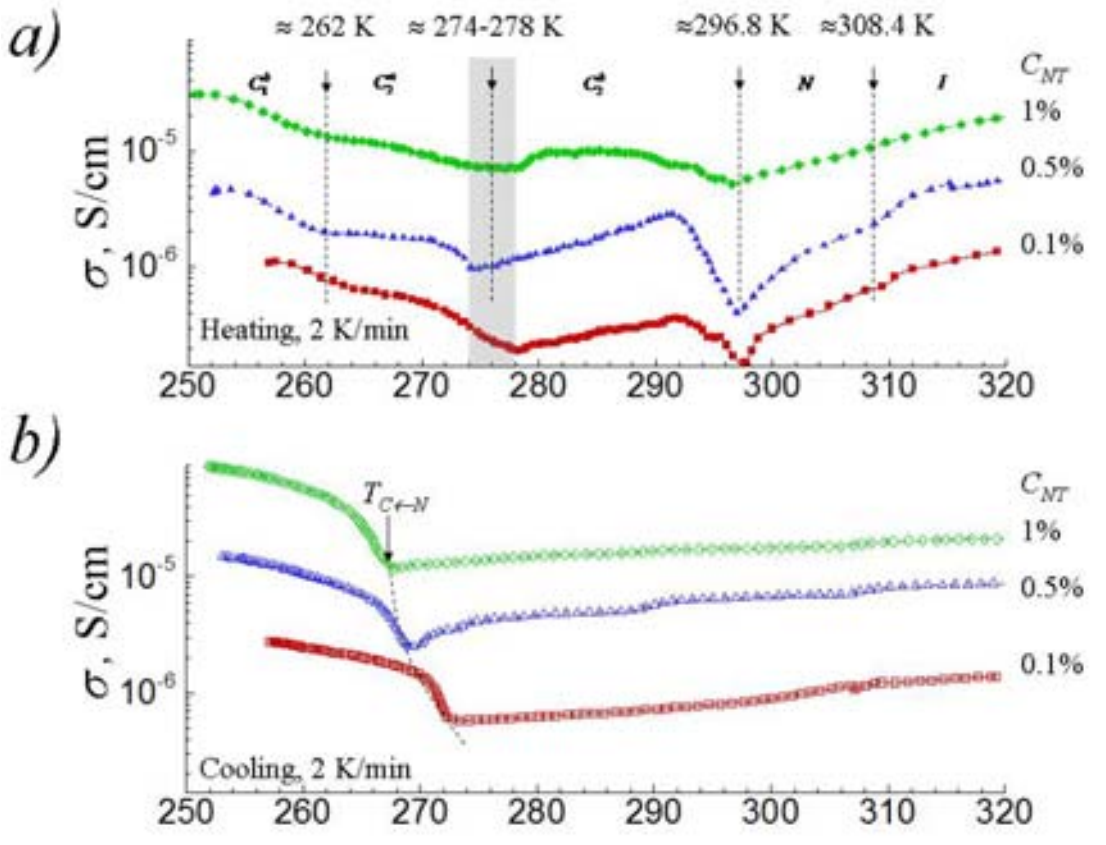

Fig. 5. Temperature dependences of electrical conductivity $\sigma$ of NTs-5CB composites for different concentrations of NTs in the heating (a) and cooling (b) regimes.

The temperature dependences of electrical conductivity $\sigma$ at different concentrations of NTs are shown in Fig.5 in heating (a) and cooling (b) regimes. In measurements of these dependences, the samples were first cooled to the solid state at 250 K, then heated to the isotropic state at 320 K and then cooled again to 250 K.

The temperature dependencies of electrical conductivity $\sigma$ in the cooling regime were typical for supercooled nematic phases and the value of $\sigma$ was noticeably increasing with commencement of crystallization[12] (Fig. 5b). The temperature dependencies of electrical conductivity $\sigma$ in the heating regime displayed several anomalies, most distinct at $T$ ≈262 K, $T$ ≈274-278 K and $T$ ≈296.8 K (Fig. 5a). The anomaly, corresponding to the melting point of 5CB ($T_{N\to I}$≈296.8 K), was assigned to manifestation of the effect of positive temperature coefficient (PTC) of resistivity. It was previously explained by the influence of thermal expansion of the LC matrix on destruction of percolation clusters, produced by multiply connected NTs [16]. The less pronounced PTC effect was observed also for transition between $C_1$ and $C_2$ crystalline phases at $T_{C_1\to C_2}$ ≈262 K. The minimums of $\sigma$ ($T$), observed in the temperature range ≈274-278 K, evidenced more complex phase behaviour inside the previously identified crystalline phase $C_2$. It fact, this phase can be subdivided into two crystalline phases: $C_2^a$ and $C_2^b$ (Fig. 5a). Behaviour of electrical conductivity was quite different inside these phases. It decreased inside $C_2^a$ phase and passed through the maximum inside $C_2^b$ phase with temperature increase. The nature of such behaviour is still unclear. However, the crystalline phases $C_2^a$ and $C_2^b$ were not observed in DSC (Fig. 2b) and photophorescence (Fig. 4) measurements that are more sensitive to the bulk properties of composites. So, we can speculate that these phases reflect the influence of phase transformations of 5CB in the interfacial layers on the transport of charge carriers between NTs.

## Conclusions

The typical nematic liquid crystal 5CB reveals a variety of phase transformations in the solid state. The previous investigations revealed the low-temperature crystal-crystal transformations at $T$≈80 K, $T$≈140 K (photoluminescence data [25]) and at $T$≈257 K (DSC data [18]). Our experimental data, obtained by DSC and photoluminescence measurements in the heating mode, evidenced the presence of two crystallization processes at $T$≈229 K and $T$≈262 K, which correspond to the phase transformations $C_1^a \rightarrow C_1^b$, and $C_1^b \rightarrow C_2$. The amplitude of these transformations may reflect the cooling pre-history of the sample. Increase of temperature $T$ from 10 K до 229 K provoked the red shift of photoluminescence spectral band that can be explained by flattening of the conformation of 5CB molecules. Crystallisation at $T$≈229 K resulted in conformation transition to non-planar 5CB structure, characteristic to ideal crystal. However, the fraction of molecules in non-planar conformation became again dominating in nematic phase, i.e. at $T$>297 K. Finally, the low-temperature phase transformations in pure 5CB may be represented by the following row: $C_o^a$ (≈80 K) $\rightarrow C_o^b$(≈140 K) $\rightarrow C_1^a$ (≈229 K) $\rightarrow C_1^b$(≈262 K) $\rightarrow C_2$ (≈296.8 K) $\rightarrow N$ (308.4) $\rightarrow I$. The introduction of NTs resulted in partial elimination of the metastable states of 5CB (e.g., $C_1^a, C_1^b$ phases). Electrical conductivity data for NTs-5CB composites revealed a supplementary anomaly inside the previously identified crystalline phase $C_2$ [18]. This phase can be subdivided in two crystalline phases: $C_2^a$ and $C_2^b$ (at $T$≈274-278 K). It can reflect the influence of phase transformations of 5CB in the interfacial layers on the transport of charge carriers between NTs.

## Acknowledgments

This work was partially funded under the projects 2.16.1.4, 65/13-H, V-153 and VC-157 (NAS of Ukraine).

## References

[1] J. P. F. Lagerwall, G. Scalia, A new era for liquid crystal research: Applications of liquid crystals in soft matter nano-, bio- and microtechnology, Current Applied Physics 12(6) (2012) 1387–1412.

[2] H. Stark, Physics of colloidal dispersions in nematic liquid crystals, Physics Reports 351 (2001) 387–474.

[3] O. Trushkevych, F. Golden, M. Pivnenko, H. Xu, N. Collings, W. A. Crossland, S. Muller, R. Jakoby, Dielectric anisotropy of nematic liquid crystals loaded with carbon nanotubes in microwave range, Electronics Letters 46(10) (2010) 693–695.

[4] C.-Y. Huang, C.-Y. Hu, H.-C. Pan, K.-Y. Lo, Electrooptical responses of carbon nanotube-doped liquid crystal devices, Japanese Journal of Applied Physics 44 (2005) 8077–8081.

[5] S. Y. Jeon, S. H. Shin, S. J. Jeong, S. H. Lee, S. H. Jeong, Y. H. Lee, H. C. Choi, K. J. Kim, Effects of carbon nanotubes on electro-optical characteristics of liquid crystal cell driven by in-plane field, Applied Physics Letters 90(12) (2007) 121901.

[6] W. Lee, H. Y. Chen, Y.-C. Shih, Reduced dc offset and faster dynamic response in a carbon-nanotube-impregnated liquid-crystal display, Journal of the Society for Information Display 16(7), (2008) 733–741.

[7] P. Malik, A. Chaudhary, R. Mehra, K. K. Raina, Electro-optic, thermo-optic and dielectric responses of multiwalled carbon nanotube doped ferroelectric liquid crystal thin films, Journal of Molecular Liquids 165 (2012) 7–11.

[8] R. Basu and G. S. Iannacchione, Carbon nanotube dispersed liquid crystal: A nano electromechanical system, Applied Physics Letters 93(18) (2008) 183105 (3 pages).

[9] S. J. Jeong, K. A. Park, S. H. Jeong, H. J. Jeong, K. H. An, C. W. Nah, D. Pribat, S. H. Lee, Y. H. S. Lee, Electroactive superelongation of carbon nanotube aggregates in liquid crystal medium, Nano Letters 7(8) (2007) 2178–2182.

[10] P. Sureshkumar, A. K. Srivastava, S. J. Jeong, M. Kim, E. M. Jo, S. H. Lee, Y. H. Lee, Anomalous Electrokinetic Dispersion of Carbon Nanotube Clusters in Liquid Crystal Under Electric Field, Journal of Nanoscience and Nanotechnology 9 (2009) 4741-4746.

[11] L. Dolgov, O. Kovalchuk, N. Lebovka, S. Tomylko, O. Yaroshchuk, Liquid Crystal Dispersions of Carbon Nanotubes: Dielectric, Electro-Optical and Structural Peculiarities. In: Carbon Nanotubes: Marulanda, J.M. Ed.; Vukovar, Croatia, InTech, 2010, pp 451–484.

[12] A. I. Goncharuk, Lebovka N. I., L. N. Lisetski, S. S. Minenko, Aggregation, percolation and phase transitions in nematic liquid crystal EBBA doped with carbon nanotubes, Journal of Physics D: Applied Physics 42(16) (2009) 165411.

[13] N. Lebovka, T. Dadakova, L. Lysetskiy, O. Melezhyk, G. Puchkovska, T. Gavrilko, J. Baran, M. Drozd, Phase transitions, intermolecular interactions and electrical conductivity behavior in carbon multiwalled nanotubes/nematic liquid crystal composites, Journal of Molecular Structure 887(1-3) (2008) 135–143.

[14] L. N. Lisetski, S. S. Minenko, A. P. Fedoryako, N. I. Lebovka, Dispersions of multiwalled carbon nanotubes in different nematic mesogens: The study of optical transmittance and electrical conductivity, Physica E: Low-Dimensional Systems and Nanostructures 41(3) (2009) 431–435.

[15] L. N. Lisetski, S. S. Minenko, V. V Ponevchinsky, M. S. Soskin, A. I. Goncharuk, N. I. Lebovka, Microstructure and incubation processes in composite liquid crystalline material (5CB) filled with multi walled carbon nanotubes, Materialwissenschaft und Werkstofftechnik 42(1) (2011) 5–14.

[16] V. V Ponevchinsky, A. I. Goncharuk, V. I. Vasil'ev, N. I. Lebovka, M. S. Soskin, Cluster self-organization of nanotubes in a nematic phase: The percolation behavior and appearance of optical singularities, JETP Letters 91(5) (2010) 241–244.

[17] N. I. Lebovka, L. N. Lisetski, A. I. Goncharuk, S. S. Minenko, V. V Ponevchinsky, M. S. Soskin, Phase transitions in smectogenic liquid crystal 4-butoxybenzylidene-4'-butylaniline (BBBA) doped by multiwalled carbon nanotubes, Phase Transitions xx(xx) (2013) 1–14.

[18] T. Mansare, R. Decressain, C. Gors, V. K. Dolganov, Phase transformations and dynamics of 4-cyano-4'-pentylbiphenyl (5CB) by nuclear magnetic resonance, analysis differential scanning calorimetry, wideangle x-ray diffraction analysis, Molecular Crystals and Liquid Crystals 382(1) (2002) 97–111.

[19] S. C. Jain, S. A. Agnihotry, V. G. Bhide, Solid crystalline polymorphism in M-24 Molecular Crystals and Liquid Crystals, Molecular Crystals and Liquid Crystals 88 (1982) 281–294.

[20] S.-W. Joon and D. Kang, Raman spectroscopy studies in 4-n-pentyl-4'-cyanobiphenyl nematic liquid crystal, in 21th International Liquid Crystal Conference, Keystone, Colorado, USA (2006).

[21] D. S. Kang, K. S. Kwon, S. I. Kim, M. S. Gong, S. S. Seo, T. W. Noh, S. W. Joo, Temperature-dependent Raman spectroscopic study of the nematic liquid crystal 4-n-pentyl-4'-cyanobiphenyl, Applied Spectroscopy 59 (2005) 1136–1140.

[22] G. W. Gray and A. Mosley, The raman spectra of 4-cyano- 4' -pentylbiphenyl and 4-cyano-4'-pentyl-d 11-biphenyl, Molecular Crystals and Liquid Crystals 35(1-2) (1976) 71–81.

[23] T. Ikeda, S. Kurihara, S. Tazuke, Excimer formation kinetics in liquid-crystalline alkylcyanobiphenyls, The Journal of Physical Chemistry B 94(17) (1990) 6550–6555.

[24] O. V Yaroshchuk, Y. P. Piryatinski, L. A. Dolgov, T. V Bidna, D. Enke, Fluorescence of the nematic liquid crystal 5CB in nanoporous glasses, Optics and Spectroscopy 100(3) (2006) 394–399.
[25] T. Bezrodna, V. Melnyk, V. Vorobjev, G. Puchkovska, Low-temperature photoluminescence of 5CB liquid crystal, Journal of Luminescence 130(7) (2010) 1134–1141.
[26] A. V Melezhik, Y. I. Sementsov, V. V Yanchenko, Synthesis of fine carbon nanotubes on coprecipitated metal oxide catalysts, Russian Journal of Applied Chemistry 78(6) (2005) 917–923.
[27] L. Liu, Y. Yang, Y. Zhang, A study on the electrical conductivity of multi-walled carbon nanotube aqueous solution, Physica E: Low-dimensional Systems and Nanostructures 24(3/4) (2004) 343–348.
[28] S. Sinton and A. Pines, Study of liquid crystal conformation by multiple quantum NMR: n-pentyl cyanobiphenyl, Chemical Physics Letters 76(2) (1980) 263–267.
[29] C. Amovilli, I. Cacelli, S. Campanile, G. Prampolini, Calculation of the intermolecular energy of large molecules by a fragmentation scheme: Application to the 4-n-pentyl-4'- cyanobiphenyl (5CB) dimer, The Journal of Chemical Physics 117(7) (2002) 3003–3012,.
[30] A. M. Klock, W. Rettig, J. Hofkens, M. van Damme, F. C. De Schryver, Excited state relaxation channels of liquid-crystalline cyanobiphenyls and a ring-bridged model compound. Comparison of bulk and dilute solution properties, Journal of Photochemistry and Photobiology A: Chemistry 85(1/2) (1995) 11–21.